\documentclass{elsart5p}

\usepackage{graphics}
\usepackage{graphicx}
\usepackage{amssymb}
\begin{document}

\begin{frontmatter}



\title{Magnetic phase diagrams of the Kagom\'{e} staircase compound Co$_3$V$_2$O$_8$}
%

\author[AA]{F. Yen\corauthref{Name1}},
\ead{fyen18@uh.edu}
\author[AA]{R. P. Chaudhury},
\author[AA]{E. Galstyan},
\author[AA]{B. Lorenz},
\author[AA]{Y. Q. Wang},
\author[AA]{Y. Y. Sun},
\author[AA,BB,CC]{C. W. Chu}

\address[AA]{TCSUH and Department of Physics, University of Houston, Houston, Texas 77204-5002, USA}
\address[BB]{Lawrence Berkeley National Laboratory, 1 Cyclotron Road, Berkeley, California 94720, USA}
\address[CC]{Hong Kong University of Science and Technology, Hong Kong, China}

\corauth[Name1]{Corresponding author. Tel: (713) 743-8305 fax: (713)
743-8201}

\begin{abstract}
At zero magnetic field, a series of five phase transitions occur in
Co$_3$V$_2$O$_8$. The N\'{e}el temperature, $T_N$=11.4 K, is
followed by four additional phase changes at $T_1$=8.9 K, $T_2$=7.0
K, $T_3$=6.9 K, and $T_4$=6.2 K. The different phases are
distinguished by the commensurability of the b-component of its spin
density wave vector. We investigate the stability of these various
phases under magnetic fields through dielectric constant and
magnetic susceptibility anomalies. The field-temperature phase
diagram of Co$_3$V$_2$O$_8$ is completely resolved. The complexity
of the phase diagram results from the competition of different
magnetic states with almost equal ground state energies due to
competing exchange interactions and frustration.
\end{abstract}

\begin{keyword}
Kagom\'{e} lattice; commensurate-incommensurate phase transitions;
geometric frustration; magnetic anisotropy
\PACS 64.60.-i,75.30.Gw,75.50.Ee,77.84.Bw
\end{keyword}

\end{frontmatter}



$Co_3V_2O_8$ is an interesting compound to be studied since
multiferroic behavior was found recently in $Ni_3V_2O_8$
\cite{paper1}. The inherent geometric frustration found in the
Kagom\'{e} lattice in $M_3V_2O_8$ (M=Co, Ni, Cu) result in the very
low temperature anti-ferromagnetic ordering of the M spins,
$T_N$=11.4 K and $T_N$=9.8 K for $Co_3V_2O_8$ and $Ni_3V_2O_8$,
respectively \cite{paper1,paper2}. The presence of two different M
sites along with competing interactions such as the single ion
anisotropy, nearest neighbor, next nearest neighbor, dipolar, and
Dzyaloshinskii-Moriya (DM) interactions allow fascinating magnetic
behavior at low temperatures such as a cascade of magnetic phase
transitions below $T_N$ \cite{paper3}, strong magnetic anisotropy
\cite{paper4}, strong sensitivity of the phase transitions with
respect to the orientation of an applied external magnetic field
\cite{paper1} and in the case for $Ni_3V_2O_8$, ferroelectricity.

In this report, we present the magnetic phase diagram of
$Co_3V_2O_8$ with fields directed along the $a$-, $b$-, and
$c$-axis. Due to the strong coupling between the magnetic orders and
the lattice all phase transitions are clearly indicated by anomalies
of the magnetization as well as of the dielectric constant. This
enables us to uniquely define and trace the phase boundaries
originating from $T_N$ to $T_4$ in the $H-T$ phase diagram. At zero
field, the 5 observed phase transitions are consistent with the
results reported by Chen et al. \cite{paper3}. Other recent reports
on the magnetic phase diagram of $Co_3V_2O_8$ focus more on higher
magnetic fields ($H\geq$ 0.5 Tesla) \cite{paper5,paper6} and they do
not resolve all details of the phase diagram, particularly in the
low-field range.

\begin{figure}
\begin{center}
\includegraphics[angle=-90,width=0.45\textwidth]{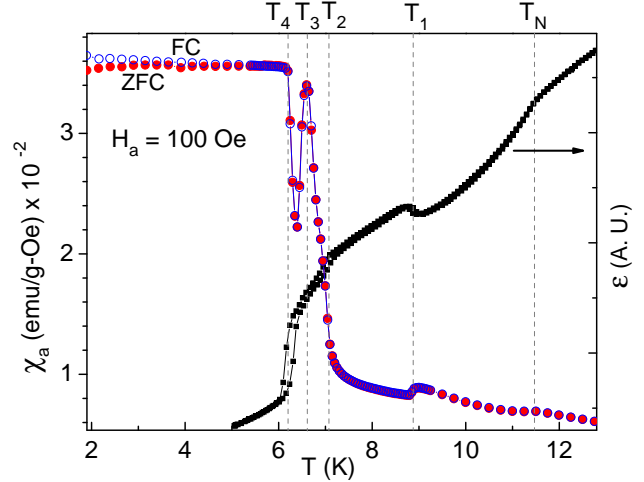}
\end{center}
\caption{Magnetization $M$ (left scale) and dielectric constant
$\varepsilon$ (right scale) of $Co_3V_2O_8$ near H=0. The five phase
transition temperatures are labeled as $T_N$ and $T_1$ to $T_4$. The
critical temperatures are clearly defined by distinct anomalies in
the $T$- and $H$-dependence of $M$ and $\varepsilon$.}
\end{figure}

In the Kagom\'{e} staircase structure of $Co_3V_2O_8$ there are two
Co$^{2+}$ sites, one located in the spines of the staircase and the
other at the cross-tie site \cite{paper3}. At $T_N$=11.4 K, only the
Co$^{2+}$ spins located in the spine site order
antiferromagnetically (AFM) along the $a$-axis. The various phases
observed in $Co_3V_2O_8$ at lower $T$ are distinguished by the
commensurability of the spin density wave vector. The $b$-component
of the magnetic modulation vector, $\delta$, decreases below $T_N$
continuously from $\delta$=0.55 and it locks-in at a commensurate
value of $\delta$=0.50 at $T_1$=8.9 K. Below $T_2$=7.0 K, $\delta$
begins to decrease continuously again locking into another
commensurate value, $\delta$=0.33, at $T_3$=6.6 K. At $T_4$=6.2 K,
$\delta$ becomes zero and the Co$^{2+}$ spins on both sites become
ferromagnetically ordered with the spin alignment along the $a$-axis
\cite{paper3}.

Single crystals of $Co_3V_2O_8$ were grown through the floating zone
furnace method. The dielectric constant was measured using Andeen
Hagerling's AH2500 capacitance bridge at a frequency of 1 kHz. The
magnetic susceptibility was measured with Quantum Design's MPMS
magnetometer.

The magnetic phase diagrams for $Co_3V_2O_8$ were constructed
through temperature and field dependent dielectric constant and dc
magnetization scans. The phase boundaries are well defined by
distinct anomalies of the magnetization as well as the dielectric
constant (Fig. 1). In order to determine precisely the stability
range of the various phases measurements of $M(T)$
($\varepsilon(T)$) as well as $M(H)$ ($\varepsilon(H)$) have been
conducted. This enabled us to resolve the details of the phase
diagram in the low-field range, for $H<$0.1 T.

For magnetic fields applied along the $a$-axis (Fig. 2a), the two
commensurate phases CM1 ($\delta$=1/2) and CM2 ($\delta$=1/3) are
stable only to about 0.07 T. Cooling below this critical field
results in a re-entrant behavior of the IC-AFM phase. Two
tricritical points exist at $T_{C1}$=6.5 K, $H_{C1}$=0.035 T and at
$T_{C2}$=9.8 K, $H_{C2}$=0.36 T where three phases coexist in a
point of the $H-T$ phase diagram. This phase diagram is different
from that derived recently from neutron scattering experiments
\cite{paper6}. The latter work resolves only three magnetic
transitions at $H$=0 due to the limited temperature and field
resolution of the experiment. Our zero-field data, however, are
consistent with the neutron work of Ref. \cite{paper3} and the
magnetic and heat capacity data of Ref. \cite{paper2}.

\begin{figure}
\begin{center}
\includegraphics[angle=0,width=0.45\textwidth]{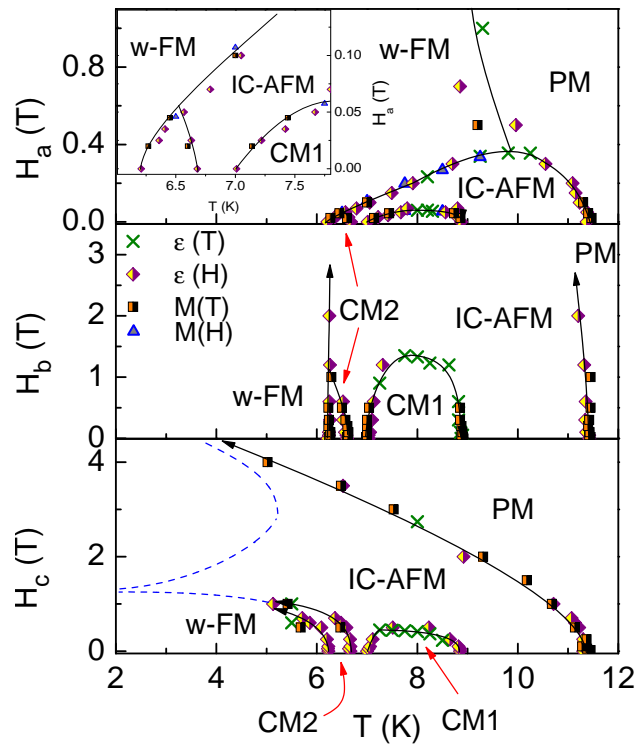}
\end{center}
\caption{Magnetic Phase diagram of $Co_3V_2O_8$ along the three
crystallographic orientations. $\epsilon$(H,T) and M(H,T) denote
dielectric constant and magnetization at varying field or
temperature. The five phases distinguished in our measurements are:
paramagnetic (PM); incommensurate AFM (IC-AFM); first commensurate
phase with $\delta$=1/2 (CM1); second commensurate phase with
$\delta$=1/3 (CM2); weak ferromagnetic phase (w-FM). The inset in
the $H_a$ phase diagram shows the details at lower magnetic fields.
The dashed lines in the $H_c$ phase diagram are extrapolations
according to Ref. \cite{paper6}.}
\end{figure}

Similar behavior is found in the $H_b$ phase diagram (Fig. 2b)
except that the stability range of the CM1 and CM2 phases extend to
about 1.2 T. Similarly, the IC-AFM phase is stable towards higher
fields ($H_b >$ 3 T) and the transition temperature into the w-FM
phase is nearly temperature independent. The first tricritical point
(coexistence of CM2, IC-AFM, and w-FW phases) is located at
$T_{C1}$=6.2 K and $H_{C1}$=1 T.

The $H_c$ phase diagram is shown in Fig. 2c. The main difference to
the phase diagrams with $H\|a$ and $H\|b$ is the stability of the
narrow CM2 phase (with $\delta$=1/3) that extends to lower
temperature at fields up to 1 T. The phase boundary at very low $T$
reported in Ref. \cite{paper5,paper6} is shown in Fig. 2c as a
dashed line.

The complex structure of the magnetic phase diagrams of $Co_3V_2O_8$
is a consequence of the existence of different magnetic states very
close in ground state energy due to the competing exchange
interactions in $Co_3V_2O_8$ and the inherent frustration. The
magnetic anisotropy, the DM and the competing nearest and next
nearest neighbor exchange interactions contribute to the richness of
the magnetic phase diagram with various commensurate and
incommensurate phases in a narrow temperature-field range. The
strong spin lattice interaction in $Co_2V_2O_8$ is evident in subtle
changes of the dielectric constant at the magnetic phase
transitions. In the sister compound, $Ni_3V_2O_8$, the
magnetoelastic coupling leads to ionic displacements and a
ferroelectric polarization in a phase with an incommensurate
transverse spiral (inversion symmetry breaking) magnetic modulation.
This transverse spiral does not exist in $Co_3V_2O_8$ \cite{paper6},
hence all magnetic phases are paraelectric.

\bibliographystyle{phpf}

\end{document}